# Layered Babinet complementary patterns acting as asymmetric negative index metamaterial

*Emese Tóth* [1, 2], *Olivér A. Fekete* [1, 2], *Balázs Bánhelyi* [1, 3] *and Mária Csete* [1, 2]*

1. Department of Optics and Quantum Electronics, University of Szeged, Dóm tér 9, Szeged 6720, Hungary;
2. Wigner Research Centre for Physics, Konkoly-Thege Miklós út 29-33., Budapest, 1121, Hungary;
3. Department of Computational Optimization, University of Szeged, Árpád tér 2, Szeged 6720, Hungary;
E-mail: mcsete@physx.u-szeged.hu

**Abstract**: Azimuthal orientation and handedness dependence of the optical responses, accompanied by asymmetric transmission and asymmetric dichroism, were demonstrated on multilayers constructed with subwavelength periodic arrays of Babinet complementary miniarrays, illuminated by linearly and circularly polarized light. In case of single-sided illumination asymmetric optical responses were observed at the spectral location of maximal cross-polarization that is accompanied by radiative electric dipoles on the nano-objects; whereas negative index material phenomenon was demonstrated, where the electric and magnetic dipoles overlap both spatially and spectrally. The NIM is accompanied by electric multipoles that add up non-radiatively and correlates with the magnetic dipoles characteristic on the nano-entities. By illuminating the multilayer with two counter-propagating circularly polarized beams it was proven that asymmetrical normal component displacement currents at the bounding interfaces arising along flat and tilted bands accompany the asymmetric co-polarized and cross-polarized transmission, respectively. The latter correlates with the asymmetric dichroism in the cross-polarized signal observed in case of single-sided circularly polarized light illumination. The dispersion maps in the single-sided asymmetrical co-polarized reflectance and absorptance indicate flat bands of analogous and complementary extrema, proving the dichroic nature of the observed asymmetric phenomena. The multilayer is proposed as an ultrathin NIM and nonreciprocal nanophotonic element.

# 1. Introduction

Metamaterials constructed with sub-wavelength patterns of plasmonic nano-objects allow for unique optical properties, including the negative index of refraction [1]. Negative index material (NIM) behavior is due to the spectrally overlapping electric dipoles excitable e.g. on wires and magnetic dipoles excitable on e.g. split ring resonators (SRR), that result in negative permittivity and permeability, respectively [2]. This explains, why primary attempts in metamaterial designs were focused on the creation of artificial magnetism via electric dipoles on circular arrays of nanoparticles or via circulating currents on SRRs [3]. Based on conformal mapping it was predicted that SRRs, acting as transformation optics analogue of infinite waveguides and supporting degenerate electric and magnetic dipoles, can result in NIM behavior in wide optical wavelength region [4].

Metamaterials are important tools in the creation of artificial chirality that can manifest itself in optical activity (OA), which is polarization rotation, and circular dichroism (CD), which is polarization selective absorption, that is comparable or outperforms those achievable via natural materials [5]. This can be achieved via e.g. three-dimensional arrays of twisted plasmonic antennas [6]. Three-dimensional chiral arrays of spherical SRRs can be used also for information encoding [7].

Asymmetric transmission (AT) on single layers was attributed to different polarization conversion efficiencies along forward and backward propagation directions for linearly or circularly polarized waves (the latter was nominated as "circular conversion dichroism"). In 2D anisotropic and lossy chiral metamaterials well-defined AT was achieved in the cross-polarized component, and it was shown that the enantiomers exhibit analogous effect with the same amplitude but reversal sign [8, 9]. Coexistent asymmetric reflection (AR) and absorption (AA) was also observed on planar arrays of asymmetric SRRs [10, 11]. AT initialization via tilting along with extrinsic OA and CD was also demonstrated [12]. The theoretical AT limit (25%) was approached in an optical spectral interval that was gradually broadened by increasing the number of circles in a spiral-shaped meta-atom [13].

Considerable asymmetric transmission in the cross-polarized state is accessible via bilayers of chiral metamaterials. Meta-atoms ensuring complete symmetry breaking arranged in bilayers can result in AT in any polarization base [14]. Twisted Babinet pattern of nanoslits and nanorods allows for AT and CD effects simultaneously [15]. Via orthogonally chained S-shaped resonator patterns with broken-symmetry along the propagation direction dual band dichroic AT effect was demonstrated in the considerable cross-polarization conversion [16].

Trilayer of bianisotropic metasurfaces constructed with patterned gold films at sub-wavelength distances allow for extremely large AT and coexistent extinction ratio at the telecommunication wavelength [17]. Circularly polarized light illumination of three layers of gold nanorods acting as anisotropic metamaterials resulted in dual-band AT with mutual polarization conversion [18].

Non-reciprocity was achieved via traveling-wave resonant ring-shaped resonators acting as magnetless nonreciprocal metamaterials coupled to microstripe-lines that can be used as circulators and isolators [19]. The key role of out-of-plane surface currents in the achievement of non-reciprocity was identified [20]. Multilayer Faraday isolator based on a resonant multi-layer cavity constructed with dichroic nanolayers and a magnetic layer was demonstrated [21]. Unidirectional as well as non-reciprocal responses were demonstrated on layered and strongly-coupled non-Hermitian plasmonic waveguides [22]. It was proven, that chirality can be combined with PT-symmetry, and may result in advanced polarization control [23].

To achieve these unique properties metamaterials consisting of plasmonic nano-objects are promising, due to the phenomena that can result in coupled modes on miniarrays [24], strong-coupling of localized and propagating modes on hole-arrays in metal films [25]. The Babinet complementary patterns are exceptionally interesting, as they exhibit complementary near- and far-field responses, as well as effective optical properties [26, 27]. Such structure can be fabricated via integrated lithography described in our previous papers [28, 29, 30, 31].

In our previous studies we have shown that a multilayer of complex patterns in convex-concave-convex succession constructed with miniarrays of spherical nano-objects results in azimuthal orientation (handedness) dependence, asymmetric transmission (as well as asymmetric dichroism (AD)) and NIM in case of illumination with linearly (circularly) polarized light [32]. Similar multilayer of complex patterns in concave-convex-concave succession was the subject of present study in order to demonstrate that the AT is accompanied by radiative electric dipoles and is due to asymmetry of displacement currents, whereas the NIM appearance is accompanied by non-radiative multipoles and is due to the overlapping electric and magnetic dipoles.

## 2. Method

The inspected multilayer is constructed with three succeeding layers of 300 nm periodic Babinet complementary patterns consisting of miniarrays of spherical nano-objects in concave-convex-concave succession. In the concave boundary layers the nano-objects are spherical holes in 39 *nm* thick metal films, whereas in the convex middle layer the nano-objects are convex gold nano-antennas of the same shape.

Namely, the central nano-object is a nanoring with R=25 *nm* outer and r=5 *nm* inner radius, this is surrounded by a nanocrescent quadrumer, the composing nanocrescents are created as an overlap of two cylinders with 0.4/5R radius. The interlayer distance is 5 *nm* in accordance with the previously inspected multilayer of convex-concave-convex patterns [32].

This pattern was illuminated by linearly and circularly polarized light in COMSOL Multiphysics (insets in **Figure 1 and Figure 2**). The complete optical responses were determined by polarization unspecific power-outflow read-out (Figure 1 a and 2 a).

The polarization specific read-out was used to determine the co-polarized and cross-polarized linearly polarized components (Figure 1 b and Figure 2 b). The co-polarized and cross-polarized circularly polarized components were determined via Jones matrix elements [32, 33]. The azimuthal orientation and handedness dependence was inspected in case of linearly and circularly polarized light, respectively. The asymmetric transmission was considered as the propagation direction dependence for the same original polarization for co-polarized and cross-polarized components selectively [16]. In case of circularly polarized light the asymmetric dichroism qualifying the direction dependence of the conversion into the orthogonal polarization states was analyzed as well. To determine the effective optical parameters and the chirality parameter of the slabs, the retrieval method relaying on the closed form Kramers-Kronig relations was used [34, 35].

The dispersion characteristic was also studied by mapping the rectified transmittance, absorptance and reflectance above the wavelength and tilting (wavenumber) parameter space, in order to identify the coupled plasmonic modes **(Figure 3 and Figure 4)**. The charge distribution at the characteristic extrema was also inspected to determine the predominant modes **(Figure 5 and Figure 6)**. The D displacement current was integrated throughout spherical shells embedding each individual nano-objects, and the supported pm magnetic dipoles were determined as the curl of the integrals. Then dynamics corresponding to the central nanoring, the nanocrescents quadrumer and complete miniarray was analyzed, by inspecting the amplitude ($p_{amp}$), projection ($p_{xy}$), inclination ($\varphi$) as well as the azimuthal orientation ($\alpha$) of the magnetic dipole (**p**$_m$) (Figure 7 and Figure 8). This study was performed at the extremum corresponding to NIM, as AT was not accompanied by magnetic dipoles in the Babinet complementary multilayer [32].

Finally, the multilayer was illuminated with counter-propagating circularly polarized light to determine the origin of asymmetric and non-reciprocal phenomena **(Figure 9 and Figure 10)**. The polarization unspecific power-outflow was also read-out for double RCP and LCP illumination (Figure 9 a and Figure 10 a).

The relative difference of the displacement current components normal to the surface $2*(D_{norm}^{top}-D_{norm}^{bottom})/(D_{norm}^{top}+D_{norm}^{bottom})$ (Figure 9 b, c and Figure 10 b, c) were compared to the $T_{++}^{forward}-T_{++}^{backward}$ and $T_{--}^{forward}-T_{--}^{backward}$ co-polarized asymmetric transmission as well as to the $T_{++}^{forward}-T_{--}^{backward}$ and $T_{--}^{forward}-T_{++}^{backward}$ co-polarized asymmetric dichroism (Figure 9 d-f and Figure 10 d-f), to the AA and AR signals (Figure 9 g, h and Figure 10 j, k), finally to the $T_{+-}^{forward}-T_{+-}^{backward}$ and $T_{-+}^{forward}-T_{-+}^{backward}$ cross-polarized AT as well as to the $T_{+-}^{forward}-T_{-+}^{backward}$ and $T_{-+}^{forward}-T_{+-}^{backward}$ cross-polarized AD signals (Figure 9 i, l and Figure 10 i, l) determined by single-sided illumination.

## 3. Result
### 3.1. Extrema in optical responses

Considering that the plasmonic resonances are accompanied by absorptance extrema, the rectified absorptance was inspected first, then the rectified transmittance and reflectance determined by polarization unspecific power-outflow read-out were compared, both for linearly and circularly polarized light illumination (Figure 1 a and Figure 2 a).

In the rectified absorptance there are well-defined maxima corresponding to the U resonance on the convex /concave pattern and to the mixed C1-C2 resonance on the concave /convex pattern in the 0° /90° azimuthal orientation (570 *nm* /560 *nm*-600 *nm*). These are followed by tiny peak /shoulder related to the C1 resonance on the concave /convex pattern and to the SPP coupled on the concave pattern (640 *nm*). Novel extrema appear, due to the interlayer coupling at 680 *nm* and 730 *nm* /720 *nm*-750 *nm*. These rectified absorptance extrema are accompanied by slightly shifted rectified transmittance maxima. An important difference is that a global maximum and a shoulder appears at 690 *nm* and 730 *nm* in the 0° azimuthal orientation, while local maxima develop at 680 *nm* and 730 *nm* in the 90° azimuthal orientation (Figure 1 a).

When circularly polarized light illumination is applied, the fingerprints of the analogous rectified absorptance maxima are observable that are coincident independent of handedness. Namely, a global maximum and shoulders appear at 570-610-640 *nm*, which are followed by larger and smaller local maxima at 680 *nm* and 730 *nm*. The novel extrema manifest themselves in the global rectified transmittance maximum at 690 *nm* and in the shoulder at 730 *nm* (Figure 2 a). The rectified reflectance is complementary to the rectified absorptance with inverted extrema at the same spectral locations, both for linearly and circularly polarized illumination (Figure 1 a, 2 a).

*3.1.1. Polarization-specific inspection of linearly polarized light transmittance*

When the multilayer is illuminated by linearly polarized light, the transmission exhibits well-defined azimuthal orientation dependence, but this can be attributed dominantly to the co-polarized light component, similarly to the convex-concave-convex multilayer [32]. In the co-polarized transmission taken in the 0° azimuthal orientation ($T_y$) shoulders appear in the interval of the single layer related resonance extrema (570 *nm* and 640 *nm*), which are followed by a well-defined local maximum and a shoulder (690 *nm* and 720 *nm*), before the global minimum observable at 750 *nm*. In comparison, in the co-polarized transmission read-out in the 90° azimuthal orientation ($T_x$) a local maximum and a shoulder (610 *nm* and 640 *nm*) correspond to the single layer related extrema, which are followed by a shoulder and a local maximum (680 *nm* and 730 *nm*), before the global minimum (760 *nm*) is reached. The maximal azimuthal orientation related difference is $8.36*10^{-3}$ / $8.47*10^{-3}$ in forward /backward propagation direction that is reached at 690 *nm* (Figure 1 b, c).

In contrast, the cross-polarized signal is tiny and shows only small local maxima preceding (550 *nm* and 620 *nm*) and succeeding (760 *nm*) the global maximum, appearing at 680 *nm* azimuthal orientation independently. Accordingly, the maximal azimuthal orientation related difference in the cross-polarized signals is $2.978*10^{-5}$ / $2.032*10^{-5}$ that is reached at 680 *nm* in forward /backward propagation direction. Both the cross-polarized signals and the maximal difference in the signals collected in orthogonal azimuthal orientations are two-orders of magnitude smaller compared to the co-polarized signals. The global maximum (680 *nm*) - minimum (740 *nm*) – local maximum (760 *nm*) in the cross-polarized signal corresponds to local maximum /shoulder (690 *nm* /680 *nm*) - shoulder /local maximum (720 *nm* /730 *nm*) – global minimum (750 *nm* /760 *nm*) in the co-polarized signal in the 0° /90° azimuthal orientation. The global minimum in the cross-polarized signal is accompanied by a shoulder / local maximum in the co-polarized transmittance (Figure 1 b, c).

Although, the forward and backward propagating signals overlap, there is a noticeable difference between the forward and backward propagating co-polarized components of the same original polarization. The maximal asymmetric transmission of $1.521*10^{-4}$ / $2.248*10^{-4}$ is reached at 680 *nm* /640 *nm* in the 0° /90° azimuthal orientation. In contrast, there is only a tiny difference between the forward and backward propagating cross-polarized transmitted light components. Namely, $2.547*10^{-5}$ / $1.325*10^{-5}$ is reached at 690 nm /690 nm in the 0° / 90° azimuthal orientation, proving negligible asymmetric cross-polarized transmission. The maximal difference is one order of magnitude smaller compared to that determined in case of the co-polarized signals **(Figure 1 b, d)**.

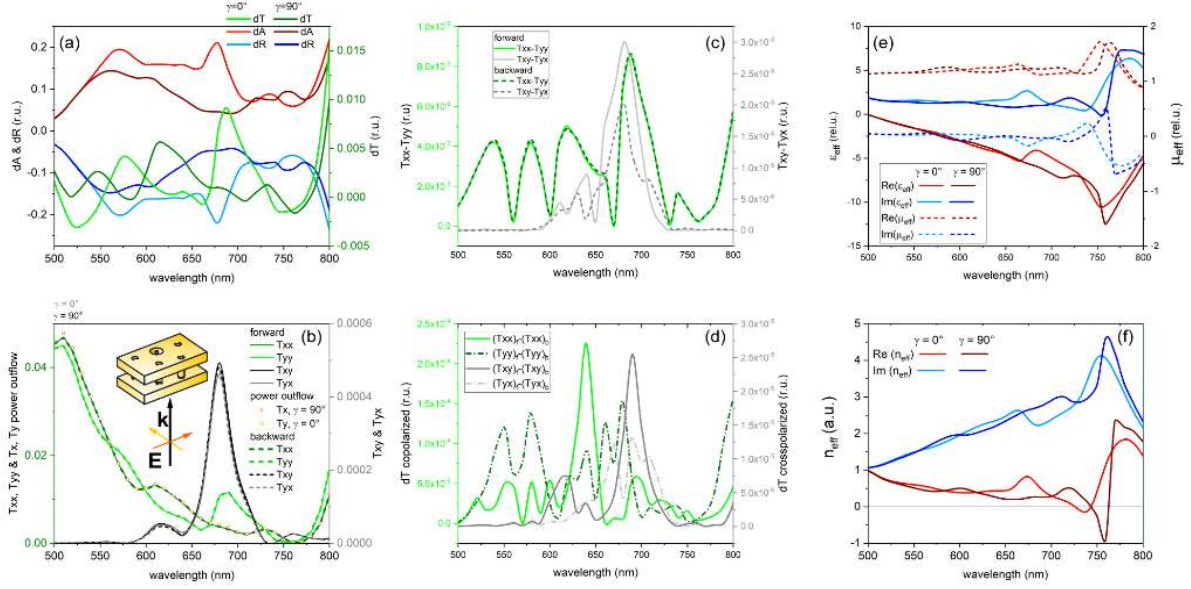

**Figure 1:** Optical responses and effective optical properties interrogated by illuminating the multilayer with linearly polarized light in the 0° and 90° azimuthal orientation. (a) Rectified optical responses from polarization unspecific read-out. (b) Co-polarized and cross-polarized transmittance with forward directed power-outflow. (c) Azimuthal orientation dependence. (d) Asymmetric transmission. Effective (e) permittivity and permeability, (f) refractive index.

*3.1.2. Polarization-specific inspection of circularly polarized light transmittance*

When circularly polarized light is used to illuminate the multilayer, the transmission is intermediate for both handednesses. The transmittance signals computed by using the Jones matrix elements exhibit more pronounced handedness dependence, than the transmission computed from the power-outflow read-out directly for LCP and RCP light illumination. The handedness dependence can be attributed dominantly to the co-polarized light component, similarly to the previously investigated convex-concave-convex multilayer [32], and analogously with the co-polarized component's dominance in case of linearly polarized light illumination **(Figure 2 b)**.

The $T_{++}^{forward}$ and $T_{--}^{backward}$ transmission signals are larger, and exhibit shoulders analogous to the single layer related resonance extrema (610 *nm* and 640 *nm*), which are followed by a well-defined local maximum and a shoulder (680 *nm* and 720 *nm*), before the global minimum at 760 *nm*. In comparison, the $T_{--}^{forward}$ and $T_{++}^{backward}$ transmission signals are smaller, and possess similar shoulders (610 *nm* and 640 *nm*) that correspond to single layer related extrema, which are followed by a local maximum and a shoulder (690 *nm* and 730 *nm*), before the global minimum at the same spectral location (760 *nm*).

The maximal handedness-related difference is 5.57*10$^{-3}$ / 5.38*10$^{-3}$ in forward /backward propagation direction that is reached at 680 *nm* (Figure 2 b, c).

In contrast, the cross-polarized signal is tiny and shows only small local maximum (560 *nm*) and a shoulder (620 *nm*) preceding, and a local maximum succeeding (750 *nm*) the global maximum, appearing at 680 *nm* handedness independently. Accordingly, the maximal handedness-related difference in the cross-polarized signals is 6.157*10$^{-5}$ /1.691*10$^{-5}$ at 680 nm in forward /backward propagation directions. Both the signals and the maximal handedness-related difference are two-orders of magnitude smaller compared to the co-polarized signals, similarly to the linearly polarized light illumination. The global maximum (680 *nm*) - minimum (740 *nm*) –local maximum (750 *nm*) in the cross-polarized signal corresponds to local maximum – (680 *nm* /690 *nm*) - shoulder (720 *nm* /730 *nm*)– global minimum (760 *nm*) in the co-polarized signal for RCP /LCP illumination (Figure 2b, c). Although, the forward and backward propagating signals overlap, there is a noticeable difference between the forward and backward propagating co-polarized light components that are analogous by means of reciprocity. The maximal asymmetrical dichroism of 1.808*10$^{-4}$ /1.417*10$^{-4}$ is reached at 640 *nm* /640 *nm*, when the $T_{++}^{forward}$ - $T_{--}^{backward}$ /$T_{--}^{forward}$ - $T_{++}^{backward}$ signal is analyzed. In contrast, there is only a tiny difference in the forward and backward propagating cross-polarized light components that are analogous by means of reciprocity. Namely, 2.598*10$^{-5}$ / 7.564*10$^{-5}$ is reached at 680 *nm* /680 *nm*, when the $T_{+-}^{forward}$ – $T_{-+}^{backward}$ / $T_{-+}^{forward}$ - $T_{+-}^{backward}$ signals are compared, proving almost negligible AD in the cross-polarized transmission. The maximal difference is one order of magnitude smaller compared to the AD reached in case of the co-polarized signals (Figure 2 b, d).

When the signals for the same original polarization are compared, significantly larger difference arises between the forward and backward propagating co-polarized components. This proves the capability for non-reciprocal responses. The maximal asymmetric transmission of 5.46*10$^{-3}$ /5.49*10$^{-3}$ is reached at 680 *nm* /680 *nm*, when the $T_{++}^{forward}$ – $T_{++}^{backward}$ / $T_{--}^{forward}$ – $T_{--}^{backward}$ co-polarized signal is analyzed. In contrast, there is only a tiny difference in the forward and backward propagating cross-polarized components. Namely, 2.115*10$^{-5}$ /8.755*10$^{-5}$ is reached at 650 *nm* /680 *nm*, when the $T_{+-}^{forward}$ – $T_{+-}^{backward}$ / $T_{-+}^{forward}$ - $T_{-+}^{backward}$ signals are compared, proving almost negligible cross-polarized AT. The maximal AD is two-orders of magnitude smaller compared to the AT reached in case of the co-polarized signals (Figure 2 b, d).

At the global minima, that appear at 750 *nm* /760 *nm* in the 0° /90° azimuthal orientation and at 760 nm both for RCP and LCP illumination in the co-polarized signal, there is a tiny local maximum in the cross-polarized transmittance signal. Moreover, not only the transmittance approaches zero, but also the azimuthal orientation and handedness dependence, as well as the AT and AD show close-to-zero minima both for the co-polarized and cross-polarized components (Figure 2 b-d). This predicts that the spectral location of the global minimum is exceptional.

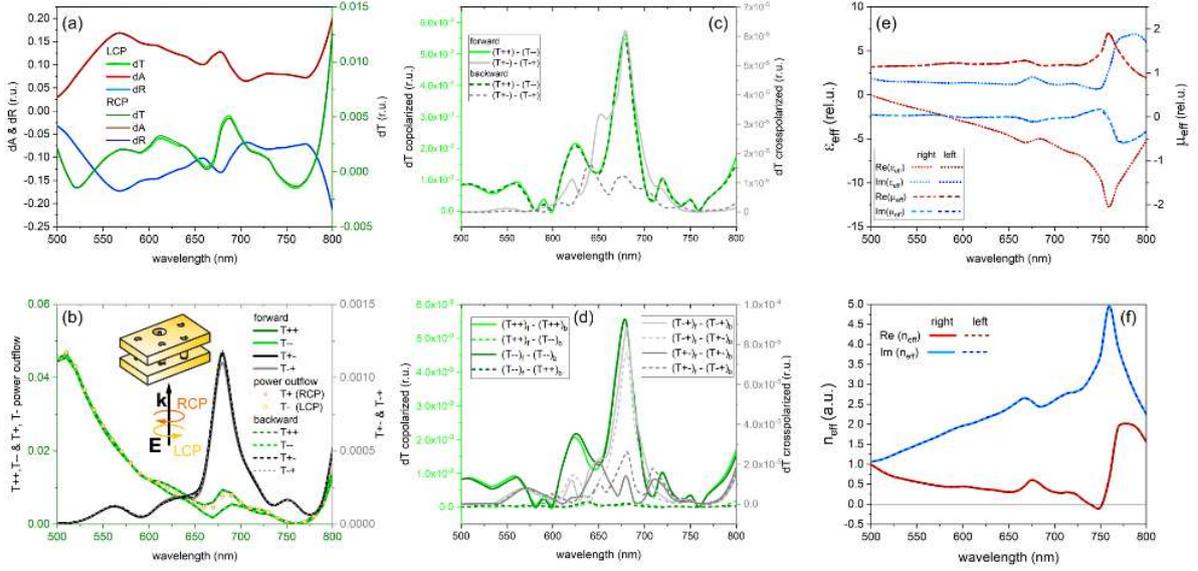

**Figure 2:** Optical responses and effective optical properties interrogated by illuminating the multilayer with right (RCP: +) and left (LCP: -) circularly polarized light. (a) Rectified optical responses from polarization unspecific read-out. (b) Co-polarized and cross-polarized transmittance with forward directed power-outflow. (c) Handedness dependence. (d) Asymmetric transmission and asymmetric dichroism. Effective (e) permittivity and permeability, (f) refractive index.

### 3.2. Effective optical parameters

The retrieval procedure resulted in NIM characteristic in all inspected configurations (Figure 1e, f and Figure 2e, f). Namely, the real part of the refractive index takes on -0.2 at 740 *nm* /-0.88 at 760 *nm*, by using linearly polarized light illumination in the 0° /90° azimuthal orientation. In comparison, the refractive index reaches a reduced value of -0.082 at 750 nm in case of illumination by circularly polarized light of either handednesses. At the same spectral location, moreover throughout the complete inspected spectral interval, the real part of the permittivity is negative, while the real part of the permeability is positive.

Lorentzian peaks appear in their imaginary parts, and the spectral location of the NIM phenomenon is almost coincident with that of the strongest magnetic resonance and the resonance peaks in the imaginary part of the index of refraction. Although, the NIM is the most pronounced in this configuration, the phase of the S21 parameter is not continuous in the 90° azimuthal orientation. Nevertheless, the Lorentzian peaks indicate spectrally overlapping electric ($\mathbf{p}_e$) and magnetic ($\mathbf{p}_m$) dipoles, which proves that they are crucial in the achievement of the NIM phenomenon (see section Magnetic dipoles on the multilayer at NIM).

In all configurations the NIM phenomenon arises at a spectral location, which is blue-shifted by 10 nm compared to the global minimum in the transmittance, except the 90° azimuthal orientation, where both the NIM and the global minimum appear at 760 nm. The azimuthal orientation dependence shows minima in the neighbor of NIM in the co-polarized /cross-polarized component at 730 *nm* /730 *nm* (740 *nm*) and 760 *nm* /770 *nm* for forward (backward) propagation. The AT takes on a minimum at 750 *nm* in all signals ($T_{yy}$, $T_{xy}$, $T_{yx}$) except the $T_{xx}$, where the global minimum appears at 760 *nm* (Figure 1 b –to – Figure 1 c, d). The handedness dependence (CD) also exhibit minima both in co-polarized and cross-polarized components, in both propagation directions at 760 *nm*. The AT and AD signals also take on a minimal value at 760 *nm* (except the $T_{+-}^{forward}$-$T_{-+}^{backward}$, which exhibits a minimum at 770 *nm*) (Figure 2 b –to – Figure 2 c, d). These coincidences prove that the NIM phenomenon appears at a spectral location, where the transmittance signal is insensitive to the azimuthal orientation and handedness, and is accompanied by vanishing asymmetric transmission and dichroism.

**3.3. Dispersion characteristic, time-evolution of the electric and magnetic modes**

*3.3.1. Dispersion characteristic: single-sided illumination*

The dispersion characteristic taken in the rectified absorptance signals shows strong /weak flat bands at 680 *nm*, and weak /strong bands at 730 *nm* /720-760 *nm* in the 0° /90° azimuthal orientation **(Figure 3 c /d)**. It is also visible that the branch corresponding the SPP coupled on the concave patterns in the (-1,0) order intersects the horizontal branch of localized modes, which results in split modes at perpendicular incidence in the 90° azimuthal orientation (Figure 3 d). The almost identical dispersion maps taken in the rectified absorptance via RCP and LCP illumination synthetize the dispersion characteristics in case of illumination by linearly polarized light of orthogonal polarizations **(Figure 4 c, d)**.

The dispersion characteristic taken in the rectified transmittance is similar, namely slightly shifted extrema appear. The global (680 *nm*) /local (640 *nm*) maximum (global maximum (680 *nm*)) at the maximal AT corresponds to an almost tilting independent flat band.

In contrast, the minimum at the NIM corresponds to an almost tilting independent side-band in case of linearly (circularly) polarized light illumination (Figure 3 a, b and Figure 4 a, b). The reflectance is complementary to these signals, the AT and NIM related bands are reversal of those in the rectified absorptance (Figure 3 e, f and Figure 4 e, f).

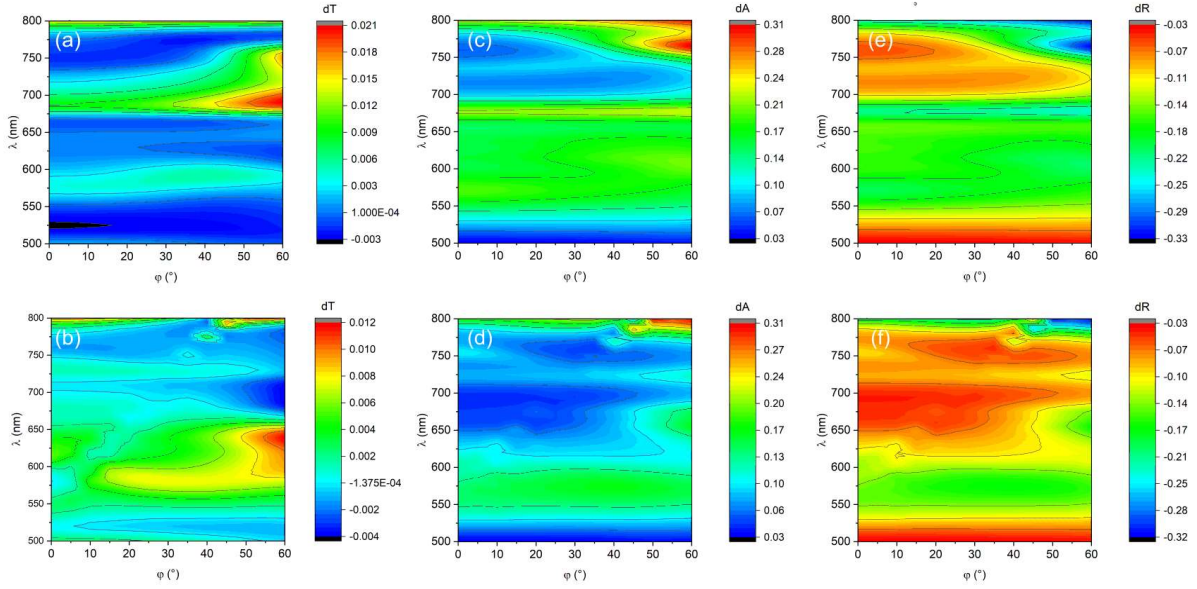

**Figure 3:** Dispersion characteristic in case of single-sided linearly polarized light illumination. Rectified (a, d) transmittance, (c, d) absorptance, (e, f) reflectance computed in the (a, c, e) 0° and (b, d, f) 90° azimuthal orientation of the multilayer.

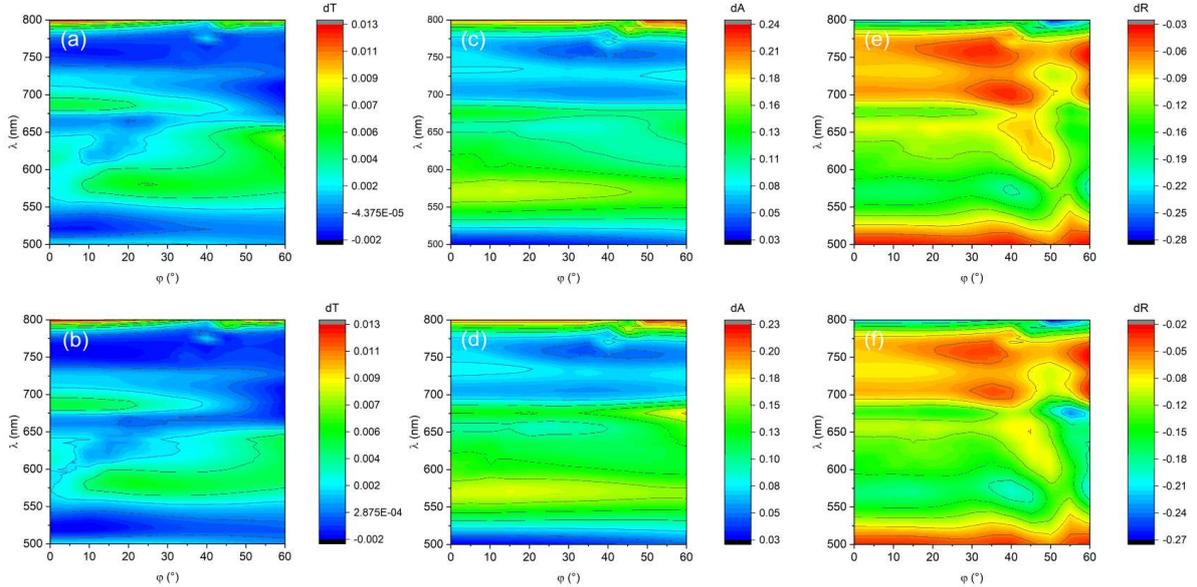

**Figure 4:** Dispersion characteristic in case of single-sided circularly polarized light illumination. Rectified (a, d) transmittance, (c, d) absorptance, (e, f) reflectance computed by using (a, c, e) RCP and (b, d, f) LCP illumination of the multilayer.

*3.3.2. Electric modes on the multilayer: AT for single-sided linearly polarized light illumination*

At the maximal AT appearing at 680 *nm* /640 *nm* in the 0° /90° azimuthal orientation of the multilayer dominantly 3D dipoles /quadrupoles develop on all nano-objects, except the central nanorings in all layers and the nanocrescents in the middle convex layer /the central nanoring in the middle convex layer, where 3D quadrupoles are also observable /dipole is dominant. Surface quadrupoles appear on the nanocrescents on the bounding metal layers in a less /more significant fraction of each time-cycle (**Figure 5** a, b, Video S1 and Video S2).

As a dominant phenomenon, the 3D dipole on the convex nanoring synchronizes the 3D quadrupoles on the concave nanorings, whereas the antiparallel 3D dipole /3D quadrupole on the convex nanocrescents synchronizes the parallel 3D dipoles /3D quadrupoles on the concave nanocrescents. The transient 3D quadrupole appearing on the convex nanoring is accompanied by reversal 3D quadrupoles /antiparallel 3D quadrupole (bottom) and dipole (top) on the concave rings; while the transient 3D quadrupoles /3D dipoles on the convex nanocrescents are accompanied by reversal 3D dipoles /dipoles and quadrupoles on the inner and outer surfaces of the concave nanocrescents.

The main difference between the two azimuthal orientations is that the surface dipoles are arranged predominantly along the y/x axis. The three-dimensional quadrupoles on the convex nanocrescents originate from vertical /in-plane charge separation. Coupled propagating SPPs also appear on the concave layers, these are intermittently in and out-of-phase on the bounding concave interfaces in the 90° azimuthal orientation.

*3.3.3. Electric modes on the multilayer: NIM for single-sided linearly polarized light illumination*

At the NIM appearing at 740 *nm* /760 *nm* in the 0° /90° azimuthal orientation of the multilayer dominantly 3D dipoles develop on all nano-objects, except the central nanoring in the middle convex/ and in the bottom concave layer as well as on each individual components of the convex /concave quadrumers, which support also 3D /outer-surface-localized lateral quadrupolar distribution intermittently (Figures 6 a, b, Video S7 and Video S8).

The succeeding bottom concave - middle convex - top concave nanorings support dominantly dipole-quadrupole-reversal dipole, and transiently quadrupole-dipole-reversal dipole. In the 0° azimuthal orientation the concave nanocrescents support dominantly parallel dipoles with an interleaved antiparallel dipole on the convex nanocrescents, and transiently antiparallel dipoles on the concave nanocrescents with interleaved quadrupoles on the convex nanocrescents.

In the 90° azimuthal orientation the nanocrescents support dominantly surface monopoles forming 3D vertical dipoles synchronized via interlayer coupling as well as quadrupoles on the complete quadrumer synchronized via inter-unit-cell coupling; while transiently lateral surface dipoles forming 3D dipoles synchronized via interlayer coupling. The main difference between the two azimuthal orientations is that the dipoles are arranged predominantly along y/x axis. In addition, interlayer coupled dipoles and inter-unit-cell coupled extended quadrupoles as well as propagating SPPs appear in the 90° azimuthal orientation.

*3.3.4. Electric modes on the multilayer: AT for single-sided circularly polarized light illumination*

At the asymmetric transmission appearing at 680 *nm* in case of RCP and LCP illumination of the multilayer predominantly 3D dipoles rotate on all nano-objects. Exceptions are the nanorings, as in the middle convex layer the nanoring supports a rotating slightly twisted dipolar distribution; while in the bottom (top) concave layer 3D quadrupole (cyclically twisted-dipole and quadrupole) rotates on the nanoring and intermittently quadrupoles also appear on the bottom concave /middle convex nanocrescents due to horizontal /vertical charge separation (Figure 5 c, d, Video S3 and Video S4).

The twisted dipole on the central convex ring mediates the charge phase that allows for rotation in [0°, 90°] interval between the nanoring surface dipoles at the interface of the bounding concave layers, while the 3D dipoles become synchronous on the nanocrescents in all succeeding layers predominantly, but are antiparallel on the convex nanocrescents both for LCP and RCP illuminations.

*3.3.5. Electric modes on the multilayer: NIM for single-sided circularly polarized light illumination*

At the NIM appearing at 750 *nm* in case of RCP and LCP illumination of the multilayer predominantly twisted 3D dipoles appear on all nano-objects. Exceptions are the middle convex nanoring, which supports permanently twisted dipolar distribution, namely perpendicular dipoles are rotating on the bottom and top surface, while in the bottom (top) concave layer strongly (slightly) twisted dipole rotates on the nanoring; and intermittently quadrupoles also appear on the uppermost and lowermost nanocrescent surfaces due to the horizontal charge separation (**Figure 6** c, d, Video S9 and Video 10). The strongly (slightly) twisted dipole on the bottom (top) concave nanoring's coupling is mediated by the central convex ring with a permanent ~60° twist that allows for [+/-120°, -/+60°] rotation between the nanoring dipoles at the interface of the bounding concave layers for RCP /LCP illumination.

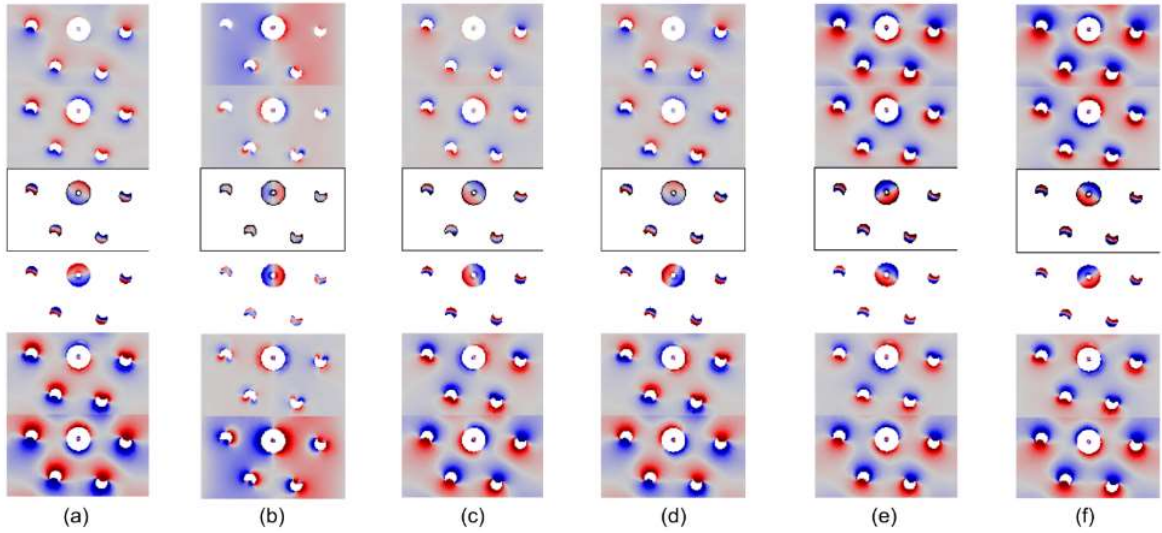

**Figure 5:** Characteristic charge distribution at the asymmetric transmission. (a/b) Linearly polarized light illumination in the 0° /90° azimuthal orientation (Video S1, Video S2). (c, e /d, f) Circularly polarized light illumination with RCP /LCP beams, from (c /d) (Video S3, Video S4) single-side and (e /f) double-side (Video S5, Video S6).

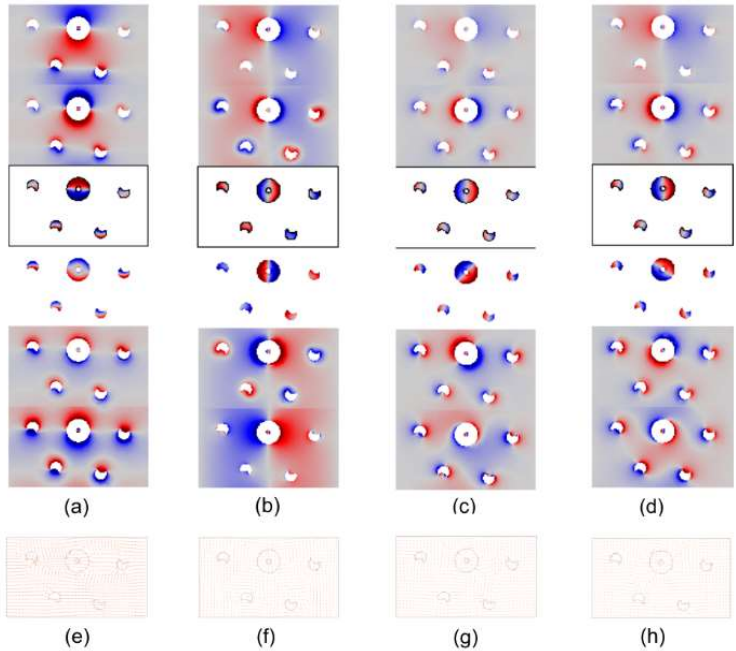

**Figure 6:** Characteristic charge distribution at the NIM and the accompanying magnetic field distribution. (a, e /b, f) Linearly polarized light illumination in the 0° /90° azimuthal orientation. (c, g /d, h) Circularly polarized light illumination with RCP /LCP beams. (a, b, c, d) charge distribution on consecutive surfaces (Video S7, Video S8, Video S9, Video S10); (e, f, g, h) magnetic field distribution on the top surface (Video S11, Video S12, Video S13, Video S14).

*3.3.6. Magnetic dipoles on the multilayer at NIM*

The magnetic dipole on the miniarray is governed by the magnetic dipole on the nanoring, according to the significantly larger amplitude and time-average of the extracted **p**$_m$. The **p**$_m$ is oriented predominantly along the x /y axis in the 0° /90° azimuthal orientation and rotates for both handedness (Figure 6 e-h, Video S11-Video S14).

Comparing the characteristic on the nanoring on the top (bottom) concave layer in the 0° /90° azimuthal orientation both the amplitude and the xy-plane projection is larger /smaller (smaller /larger); the inclination is considerably smaller /larger; the rotation is slow /quick. On the convex nanoring the amplitude and projection is smaller /larger, the inclination is considerably smaller /larger, the rotation is slow /quick. The magnetic dipole dynamics on the quadrumer of the nanocrescents on the top (bottom) layer in the 0° /90° azimuthal orientation indicates slightly smaller /larger amplitude and projection; larger reversal /smaller synchronous (smaller synchronous /larger reversal) inclination; slow (continuous) /quick rotation.

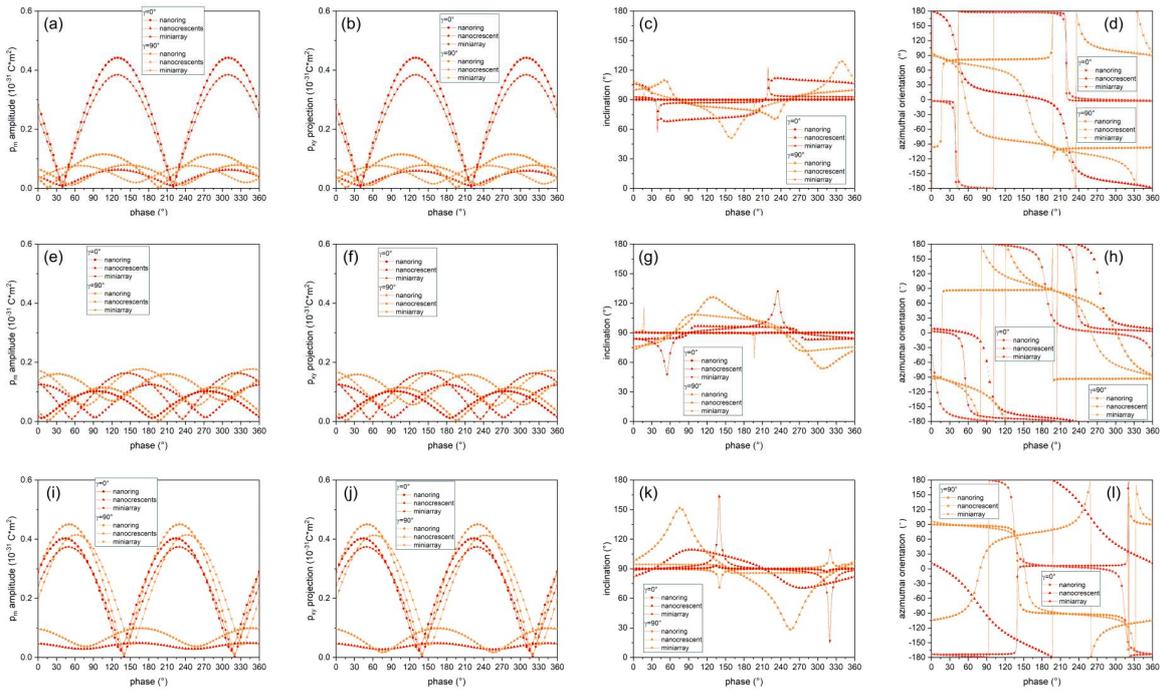

**Figure 7:** Dynamics of the accompanying magnetic dipoles in case of linearly polarized light illumination. (a, e, i) amplitude and (b, f, j) projection; (c, g, k) inclination; (d, h, l) azimuthal orientation; in the (a-d) / (i-l) top /bottom concave and (e-h) middle convex layer.

On the convex quadrumer the amplitude and projection is similarly smaller /larger, the inclination is smaller synchronous /larger reversal, and the rotation is quick /slow.

These together result in magnetic dipoles on the miniarray on the top (bottom) concave layer in the 0° /90° azimuthal orientation, that is accompanied by larger /smaller (smaller /larger) amplitude and projection; noticeably smaller /larger (larger /smaller) inclination; significantly quicker /slower rotation. On the convex miniarray the amplitude and projection is smaller /larger, the inclination is larger /smaller, while the rotation is quick /slow **(Figure 7)**.

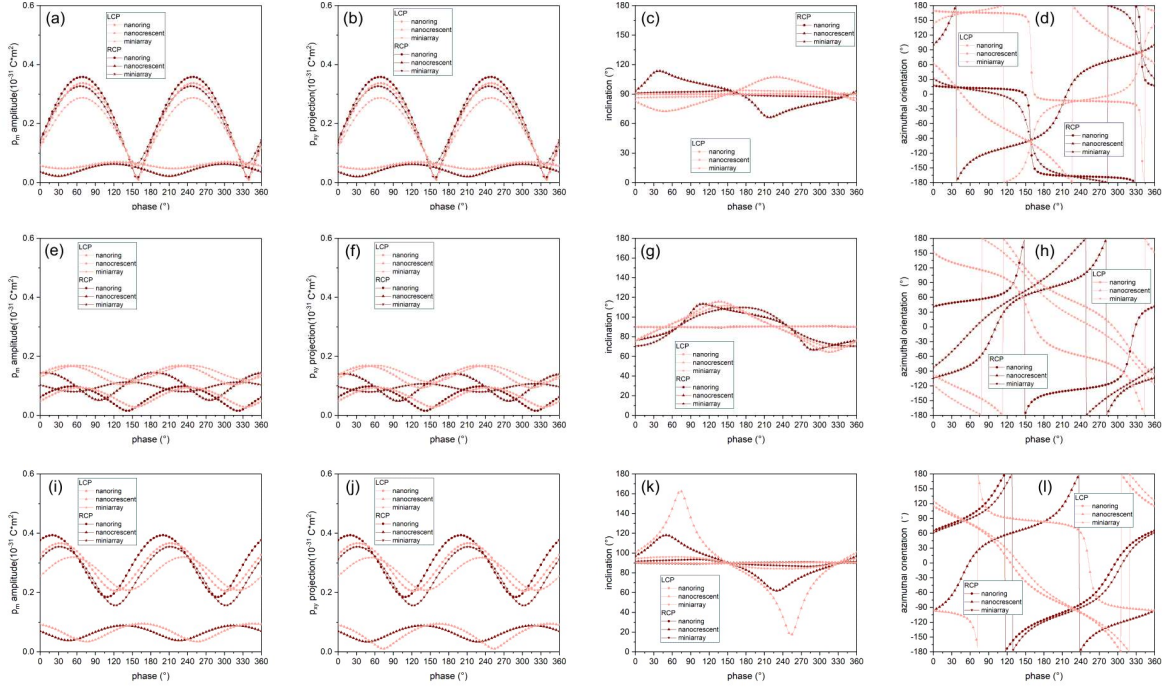

**Figure 8:** Dynamics of the accompanying magnetic dipoles in case of circularly polarized light illumination. (a, e, i) amplitude and (b, f, j) projection; (c, g, k) inclination; (d, h, l) azimuthal orientation; in the (a-d) / (i-l) top /bottom concave and (e-h) middle convex layer.

The characteristic of the convex magnetic dipoles is inherited predominantly from the bottom layer, and the 90° azimuthal orientation is slightly more advantageous. Namely, the azimuthal orientation preference for the amplitude & projection on the nanoring and miniarray, and the inclination on the quadrumer and miniarray is analogous with the bottom layer; moreover for the inclination and rotation on the nanoring, for the amplitude and projection on the quadrumer as well as for the rotation on the miniarray it is the same in all layers, while the rotation exhibits distinct azimuthal orientation dependence on the quadrumer. Comparing the characteristics on the nanoring on the top/ (bottom) layer for the RCP /LCP light illumination, both the amplitude and the xy-plane projection is slightly larger /smaller; the inclination is slightly smaller /larger (larger /smaller); the turning is (slightly) longer/shorter.

On the convex nanoring the amplitude and projection is slightly larger /smaller, the inclination is larger /smaller, while the rotation is quick /slow. The magnetic dipole dynamics on the quadrumer of the nanocrescents on the top (bottom) layer for the RCP /LCP light illumination indicates slightly smaller /larger amplitude and projection; significantly larger /smaller (smaller /larger) inclination; quick /slow (slow / quick) rotation. On the convex quadrumer the amplitude and projection is smaller /larger, the inclination is slightly smaller /larger, and the rotation is quick /slow continuous. These together result in magnetic dipoles on the miniarray on the top (bottom) layer for the RCP /LCP light illumination, that is accompanied by larger /smaller amplitude and projection; slightly smaller /larger inclination; longer/shorter (shorter /longer) turning. On the convex miniarray, the amplitude and projection is smaller /larger, the inclination is smaller /larger, the turning is longer /shorter **(Figure 8)**.

The characteristic of the convex magnetic dipoles is inherited partially from the bottom and top layers, but the handedness preference of fewer properties is analogous for the three layers. Namely, the handedness preference of the inclination on the convex nanoring and quadrumer is analogous with the bottom layer; for the amplitude & projection on the nanoring and quadrumer as well as for the inclination on the miniarray it is the same for all layers; while for the amplitude and projection on the miniarray and for the rotation on the nanoring it exhibits a distinct azimuthal orientation dependence, and the rotation is more likely to the top layer on the quadrumer and on the miniarray.

### 3.4. Optical response, dispersion characteristics, time-evolution of the electric modes: double-sided illumination

*3.4.1. Extrema in optical responses: double-sided illumination*

The rectified absorptance (dA) and the rectified outflow (d(T+R)), determined by polarization unspecific power-outflow read-out, were compared for the RCP and LCP light illuminations (**Figure 9** a and **Figure 10** a). In addition to this, the difference between the outflows in opposite propagation directions ($\Delta$(T+R)) was also analyzed. In the rectified absorptance signal there are well-defined maxima corresponding to the single layer related maxima (570 *nm* and 640 *nm*), which are followed by the extrema due to the interlayer coupling at 680 *nm* and 760 *nm*. The rectified absorptance extrema are accompanied by slightly shifted complementary rectified outflow modulations, but an important difference is that a minimum appears at 570 nm, 680 *nm*/ 670 *nm* and 760 *nm*, while a tiny local maximum develops at 650 *nm*/ 640 *nm* in case of RCP /LCP illumination (Figure 9 a and Figure 10 a).

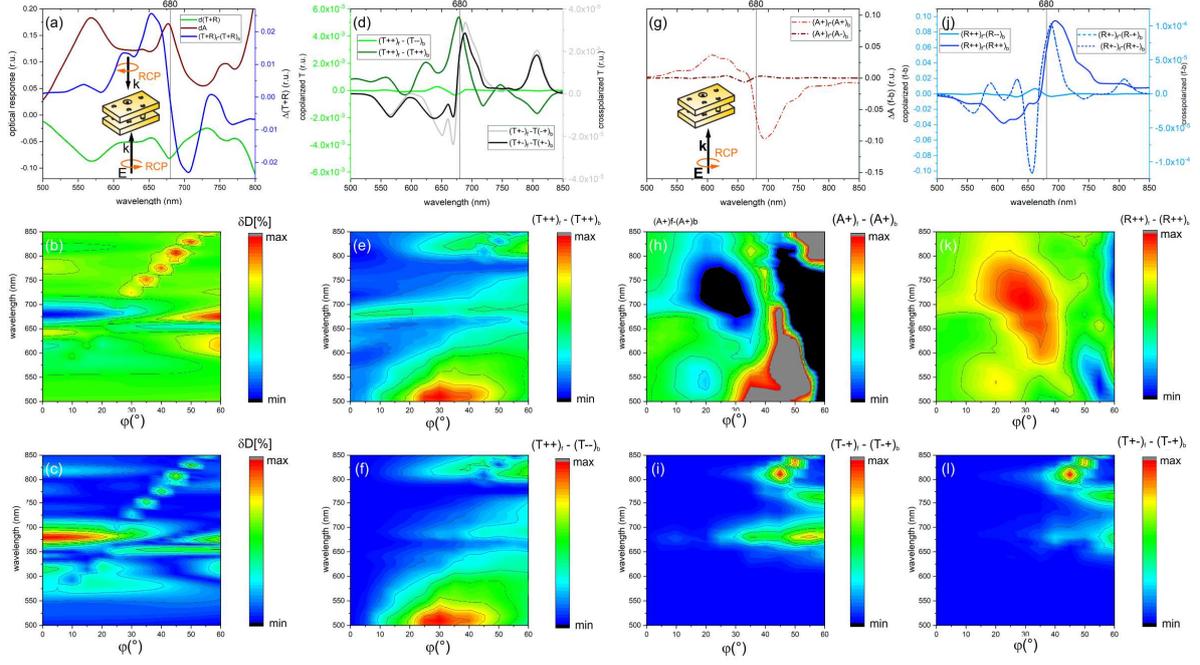

**Fig. 9:** Origin of the asymmetric transmission in case of single and double sided illumination by RCP light. (a) Polarization unspecific power-outflow; dispersion maps of the (b) value and (c) absolute value of the relative difference in the normal component displacement current in case of double-sided illumination. (d, g, j) Co-polarized and cross-polarized asymmetric signals and (e, h, k) co-polarized dispersion maps in (d, e) transmittance, (g, h) absorptance, (j, k) reflectance; and dispersion maps in the (f) asymmetric dichroism, cross-polarized asymmetric (i) transmission and (l) dichroism in case of single-sided illumination.

In case or double RCP /LCP illumination the differential outflow signal exhibits analogous reversal extrema at the single layer U and mixed C1-C2 resonance related maxima (560 *nm* and 620 *nm*), while the C1 resonance overlapping with coupled SPP related extremum appears at the same location (650 *nm*) /is forward shifted (670 *nm*) compared to the extrema from single-sided illuminations. An inflection corresponding to zero-crossing and succeeding a local maximum /minimum is observable at the AT (680 *nm*) and NIM (750 *nm*) locations observed in case of single-sided illumination.

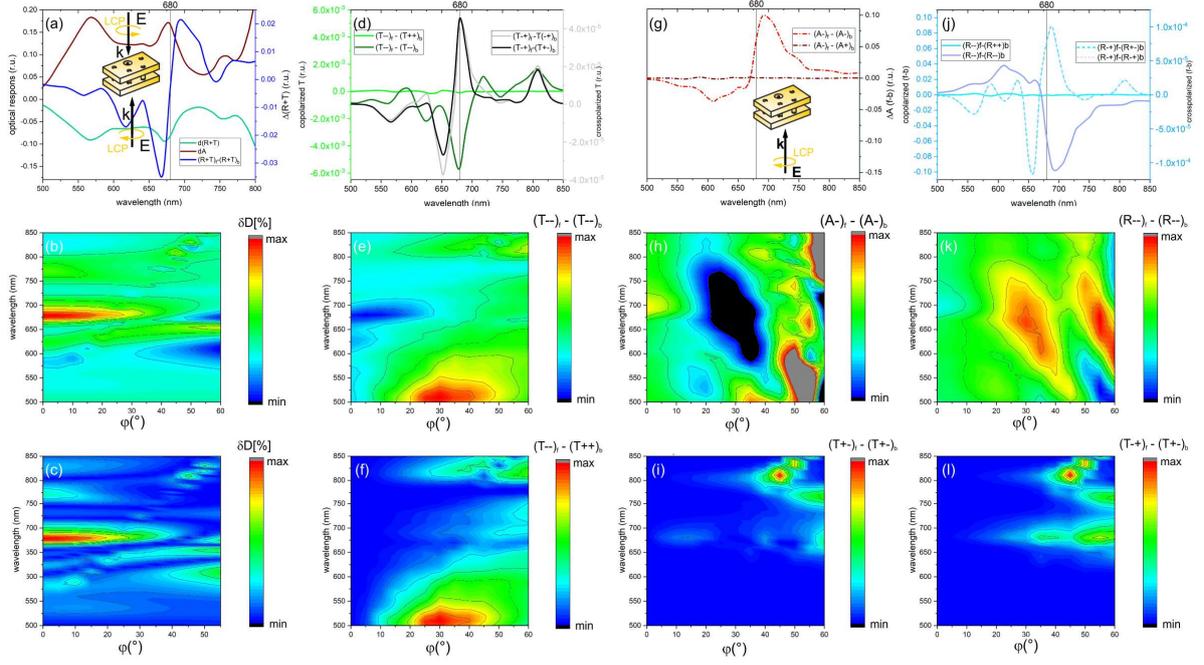

**Figure 10:** Origin of the asymmetric transmission in case of single and double sided illumination by LCP light. (a) Polarization unspecific power-outflow; dispersion maps of the (b) value and (c) absolute value of the relative difference in the normal component displacement current in case of double-sided illumination. (d, g, j) Co-polarized and cross-polarized asymmetric signals and (e, h, k) dispersion maps in (d, e) transmittance, (g, h) absorptance, (j, k) reflectance; and dispersion maps in the (f) asymmetric dichroism, cross-polarized asymmetric (i) transmission and (l) dichroism in case of single-sided illumination.

*3.4.2. Dispersion characteristic: double-sided illumination – compared to single-sided asymmetry*

According to the intuitive expectation, the dispersion map taken in the relative difference of the displacement current components normal to the surface 2* ($D_{norm}^{top}$-$D_{norm}^{bottom}$)/ ($D_{norm}^{top}$+$D_{norm}^{bottom}$) indicates a well-defined band of minima /maxima for RCP /LCP double-sided illumination at the spectral location of the AT phenomenon. This flat band is extended throughout the tilting, where crossing occurs with a tilted band that resembles the band of propagating SPPs. Above this crossing maxima appear in the $\delta D_{norm}$ along the tilted branch for both handednesses (Figure 9 b, c and Figure 10 b, c).

On the differential dispersion maps taken in the $T_{++}^{forward}$-$T_{++}^{backward}$ and $T_{--}^{forward}$-$T_{--}^{backward}$ co-polarized AT signals a well-defined flat band of maxima /minima appears at 680 *nm* through 30°/15° tilting, the extension of the band is larger /smaller in case of RCP /LCP illumination (Figure 9 d, e and Figure 10 d, e).

In contrast, on the differential dispersion maps taken in the $T_{++}^{forward}$-$T_{--}^{backward}$ and $T_{--}^{forward}$-$T_{++}^{backward}$ co-polarized AD signal the corresponding flat band of extrema has a fingerprint at large tilting rather than at perpendicular incidence (Figure 9 f and Figure 10 f). At the tilted band corresponding the propagating SPP modes, both differential co-polarized transmittance signals exhibit minima. These results prove that well-defined co-polarized bands correspond to the AT phenomenon for both handednesses and the differential $\delta D_{norm}$ signal is related unambiguously to co-polarized AT rather than to AD (Figure 9 e, f and 10 e, f). However, the co-polarized AT extrema anti-correlate with the extrema appearing along the flat-and-tilted bands in the $\delta D_{norm}$.

The characteristic of the $\delta D_{norm}$ can be better explained by considering the cross-polarized AT and AD of the signals, as on all differential dispersion maps taken in the $T_{+-}^{forward}$-$T_{+-}^{backward}$ and $T_{-+}^{forward}$-$T_{-+}^{backward}$ cross-polarized AT signals and in the $T_{+-}^{forward}$-$T_{-+}^{backward}$ and $T_{-+}^{forward}$-$T_{+-}^{backward}$ cross-polarized AD signals there are maxima along the flat and tilted bands (Figure 9 d, i, l and Figure 10 d, i, l).

However, the degree of modulation in the transmittance signals is very small, therefore it is reasonable to inspect the reflectance and transmittance of specific polarization as well. The map taken in the co-polarized reflectance indicates a flat band of maxima /minima at the AT phenomenon, while the absorptance shows a flat band of minima /maxima (Figure 9 g, h, j, k and Figure 10 g, h, j, k). This unambiguously proves that the AT is of dichroic origin, and the differential absorption is divided into two channels of asymmetric transmission and reflection.

*3.4.3. Electric modes on the multilayer: AT for double-sided circularly polarized light illumination*

In case of double RCP and LCP illuminations of the multilayer at 680 *nm* perfect 3D quadrupoles rotate clockwise (counter-clockwise) and counter-clockwise (clockwise) on the bottom (top) concave nanorings and parallel 3D dipoles appear on the concave nanocrescents; while the twisted multipole continuously varies between a dipole and a quadrupole on the convex nanoring, but the convex nanocrescents support 3D dipoles that are antiparallel compared to the parallel ones on the concave nanocrescents (Figure 5 e, f, Video S5 and Video S6).

On the concave nanorings the relative orientation of the quadrupoles continuously varies in the [0°, 180°], due to the counter-varying phases, that is mediated by the continuously varying multipole on the convex nanoring. The predominant 3D dipoles on the concave nanocrescents are parallel due to the interleaved antiparallel 3D dipole on the convex nanocrescents.

On the concave nanocrescents transiently quadrupoles appear due to horizontal charge separation, but the 3D dipoles on the convex nanocrescents quickly switch their orientation, therefore higher order multipoles cannot appear. The surface dipoles are perfectly in-phase in the quadrumer in each specific layers. In contrast, the surface multipoles on the nanocrescents exhibit a small phase delay in the quadrumers on the lower and upper surfaces of each layers.

**4. Conclusion**

The multilayer illuminated by linearly polarized light exhibited azimuthal orientation dependence and asymmetric transmission, both were one order of magnitude larger in the co-polarized component. The AT was coincident with the spectral location of the maximum in the cross-polarized signals.

The multilayer illuminated by circularly polarized light exhibited handedness and propagation direction dependence that manifested itself in asymmetric transmission and dichroism, these were one and two-orders of magnitude larger in the co-polarized components. The AT was coincident with the spectral location of the maximum in the cross-polarized signals (except the T+- cross-polarized signal), whereas the co-polarized /cross-polarized AD inherited the maximum in the co-polarized AT in the 90° /0° azimuthal orientation.

NIM was achieved both with linearly and circularly polarized light illumination at the spectral locations, where electric and magnetic dipoles overlap. However, all signals, as well as their azimuthal orientation and handedness dependence exhibit a minimum close to the NIM phenomenon. This already predicts the non-radiative nature of the multipoles that are excitable at the NIM.

The dispersion characteristics taken via single sided illumination exhibits a flat band of extrema at the AT phenomenon and a side band of inverted extrema at the NIM phenomenon, respectively.

In case of linearly polarized light illumination at the AT phenomenon the predominant dipole on the convex nanoring allows for transmission that is larger due to the accompanying synchronized nanocrescent dipoles in the 0° azimuthal orientation. In contrast, at the NIM phenomenon zero transmittance is caused by the reversal predominant dipoles on the nanorings in the bounding concave layers accompanied by partially synchronized dipoles on the nanocrescents. The large fraction of the reversal dipoles on the nanoring and nanocrescents in the top concave /as well as in the middle convex layer is not advantageous for transmission.

In case of circularly polarized light illumination the predominant twisted dipole on the convex and upper concave nanoring accompanied by well-synchronized nanocrescent dipoles allow for finite transmission at the AT phenomenon. In contrast, the zero transmittance is caused by predominant twisted dipoles on the nanorings in the bounding concave layers with continuously varying phase accompanied by less synchronized twisted dipoles on the nanocrescents at the NIM phenomenon. The larger fraction of the (perpendicular) reversal dipoles on the nanoring and nanocrescents in the (bottom) top concave and middle convex layer is also not advantageous for transmission.

At the spectral location of the AT phenomenon in case of linearly (circularly) polarized illumination the charge distribution is localized, and interlayer-coupling occurs between individual nano-objects. The transmittance maximum /shoulder (maximum) is due to the significant fraction of radiating dipoles in the 0° /90° azimuthal orientation (including the twisted dipoles on the convex and upper concave nanoring and the synchronized dipoles on the nanocrescents for both handednesses). The propagating grating-coupled SPPs are intermittently in phase, i.e. transmission is cyclically promoted in case of linearly polarized light illumination in the 90° azimuthal orientation, whereas they are permanently in phase, i.e. the transmission is promoted for both handednesses.

The NIM differs from the AT location in the sense that the charge distribution is more extended, interlayer and inter-unit cell coupling also appears due to the synchronized surface monopoles on individual nano-objects, that are capable of preventing the transmission. In addition to this, the localized modes are out-of-phase with the propagating SPPs, and the quadrupoles appearing on the complete quadrumer (in the middle convex and top concave layer) are non-radiative in the 90° azimuthal orientation (for both handednesses), thereby forward radiation, i.e. transmission is prevented.

Considering all of the three layers, the magnetic dipoles characteristic predicts similar NIM in the 0° and 90° azimuthal orientations, which correlates with that although, in the 90° azimuthal orientation the NIM phenomenon seems to be more pronounced, it is accompanied by a discontinuous phase of the S21 parameter used for computation. The balanced magnetic dipoles characteristic predicts similar NIM for the two handednesses, which correlates with the extracted effective parameters.

The double-sided illumination results in optical response modulation in the interval of single and multilayer specific extrema, but in the differential outflow signal inflections are coincident with the spectral locations of AT and NIM that were observed in case of single-sided illumination.

The dispersion maps taken via double-sided illumination in the relative difference of normal-component displacement currents and in the single-sided differential transmittance signals indicate flat and tilted bands corresponding to the co-polarized and cross-polarized AT phenomenon, respectively, that are accompanied by weak flat and similar tilted bands in the differential signals corresponding to the cross-polarized AD phenomenon. The dispersion maps taken in the single-sided differential co-polarized reflectance and absorptance indicate well-defined flat bands of analogous and complementary extrema proving the dichroic origin of the observed AT phenomena.

In case of double-sided RCP and LCP illumination at the AT location the charge distribution is localized, interlayer-coupling occurs between individual nano-objects. The outflow drop is due to the nearby absorptance related to the coexistence of radiating dipoles on the nanocrescents and convex nanoring as well as of non-radiating quadrupoles on the concave nanorings. The propagating SPPs are cyclically in phase, i.e. transmission is promoted.

The multilayer exhibits NIM and non-reciprocal characteristic as well, the users have to define, which phenomenon is preferable for a specific application.

**Conflict of interest**: The authors declare no conflicts of interest regarding this article.

**Supporting Information**

Supporting Information (Video S1 - Video S14) is available from the authors.

**Acknowledgements**

This work was supported by the National Research, Development and Innovation Office (NKFIH) of Hungary, projects: "Optimized nanoplasmonics" (K116362), "Nanoplasmonic Laser Inertial Fusion Research Laboratory" (NKFIH-2022-2.1.1-NL-2022-00002) and "National Laboratory for Cooperative Technologies" (NKFIH-2022-2.1.1-NL-2022-00012) in the framework of the Hungarian National Laboratory program.